\documentclass[12pt]{article}
\usepackage{a4wide,amsmath,amsthm,epsfig}
\usepackage{amsmath,amsthm}
\usepackage{hhline}
\usepackage{amsfonts}

\newcommand{\url}[1]{}

\newcommand{\Qx}{ \mathbb{Q} }
\newcommand{\Ex}{ \mathbb{E} }

\newcommand{\cds}{\mbox{CDS}}

\newcommand{\pcds}{\Pi\mbox{\tiny RCDS}}

\newcommand{\npv}{\mbox{NPV}}

\newcommand{\Hhat}{\widehat{H}}
\newcommand{\lgd}{\mbox{L{\tiny GD}}}
\newcommand{\rec}{\mbox{R{\tiny EC}}}

\newtheorem{theorem}{Theorem}[section]
\newtheorem{proposition}[theorem]{Proposition}
\newtheorem{remark}[theorem]{Remark}
\newtheorem{definition}[theorem]{Definition}

\title{{\small This paper is available at www.damianobrigo.it\\ First Posted at ssrn.com on March 10, 2005}\\
\vspace{1cm} {\large \bf Credit Default Swap Calibration  and
Counterparty Risk Valuation with a Scenario based First Passage
Model\thanks{We are grateful towards Eymen Errais, for suggesting
us to have a look at the random barrier approach, and to Kay
Giesecke for helpful discussion and correspondence.}}}
\author{
Damiano Brigo \ \ \ Marco Tarenghi\\
Credit Models -
Banca IMI\\
Corso Matteotti 6,
20121 Milano, Italy\\
\texttt{\{damiano.brigo, marco.tarenghi\}@bancaimi.it} \\
\texttt{http://www.damianobrigo.it}\\
}

\date{First version: March 1, 2004. This Version: April 29, 2005}

\begin{document}

\maketitle \thispagestyle{empty}

\begin{abstract}
In this work we develop a tractable structural model with
analytical default probabilities depending on a random default
barrier and possibly random volatility ideally associated with a
scenario based underlying firm debt. We show how to calibrate this
model using a chosen number of reference Credit Default Swap (CDS)
market quotes. In general this model can be seen as a possible
extension of the time-varying AT1P model in Brigo and Tarenghi
(2004). The calibration capability of the Scenario
Volatility/Barrier model (SVBAT1P), when keeping time-constant
volatility, appears inferior to the one of AT1P with time-varying
deterministic volatility. The SVBAT1P model, however, maintains
the benefits of time-homogeneity and can lead to satisfactory
calibration results, as we show in a case study where we compare
different choices on scenarios and parameters.

Similarly to AT1P, SVBAT1P is suited to pricing hybrid
equity/credit derivatives and to evaluate counterparty risk in
equity payoffs, and more generally to evaluate hybrid
credit/equity payoffs. We consider the equity return swap in Brigo
and Tarenghi (2004) and show its valuation under SVBAT1P with the
same CDS and equity calibration input used earlier for AT1P, and
further we hint at equity default swap valuation in the
conclusions.
\end{abstract}


\newpage


\noindent{\em Duke of Alencon, this was your default,\\ That,
being captain of the watch to-night,\\ Did look no better to that
weighty
charge.}\\

King Henry VI, Part 1, Act 2, Scene I.\\

\section{Introduction}

Modelling firms default is becoming more and more important,
especially in recent times where the market is experiencing a
large development in credit derivatives trading. In this paper we
develop a tractable structural model with analytical default
probabilities depending on some dynamics parameters and on a
random default barrier ideally associated with the underlying firm
debt. The scenario based default barrier models our possible
uncertainty on the debt level of the firm. This uncertainty may be
due to the fact that, for example, we do not trust completely the
information from Balance Sheets.

We show how to calibrate the model using a chosen number of Credit
Default Swap (CDS) market quotes. We apply the structural model to
a concrete calibration case. This model can be seen as a variant
or even a possible extension of the time-varying Analytically
Tractable 1st Passage model (AT1P) in Brigo and Tarenghi (2004).
There are a number of approaches for extending this earlier work.
In Brigo and Tarenghi (2004) we essentially used information from
equity volatility and CDS quotes to determine a time-varying value
of the deterministic firm-volatility and a reference level for the
default barrier implying exact calibration of a given number of
CDS quotes. In this paper we consider a scenario based default
barrier/volatility version of the basic model, which is a step
forth, but take time-constant volatilities (in each scenario),
which is a step backwards. Notwithstanding the time-constancy of
the volatility scenarios, which is a restriction with respect to
the earlier deterministic time-varying coefficients AT1P model, we
refer to the random coefficients model as to an ``extension", that
we term Scenario Volatility/Barrier Analytically Tractable 1st
Passage model (SVBAT1P). The calibration of the CDS term structure
in SVBAT1P is left to the random default barrier, to the
volatility scenarios parameters and to the probabilities of
different scenarios. These can all be considered as calibrating
parameters, or else some can be fixed arbitrarily and the
calibration can be left to the remaining ones. We illustrate
different possible choices with a case study.

With the first extension where we use scenarios only on the
barrier (taking instead the same volatility in all scenarios) the
model is not very flexible and can calibrate a low number of CDS,
typically 3 or 4. Indeed, default barrier scenarios appear not to
be natural fitting parameters for a large number of CDS quotes. In
this case it can be best to associate scenarios also to
volatilities, or to take a time varying volatility to be partially
fitted to CDS quotes even in the case of random barrier. In this
note we explore the former solution, and while time-constant
volatility scenarios do not allow a perfect fit of all the five
CDS maturities, they still allow a small calibration error, which
can be deemed to be acceptable when compared to the market induced
bid ask spreads. What is more, having a model like SVBAT1P with
constant parameters, even if under different scenarios, may help
with respect to robustness, typically endangered by all-fitting
time-varying functions as in the basic AT1P model. Nonetheless,
the time-varying volatility scenarios extension will be addressed
in further work.

Finally we notice that, as for the earlier deterministic barrier
AT1P model, the CDS calibrated SVBAT1P model is ideally suited to
price hybrid equity/credit derivatives and to evaluate
counterparty risk in equity payoffs. Given the same CDS and equity
calibration inputs, we compare the price of an equity return swap
under the deterministic time varying volatility model AT1P with
the same price under the scenarios model SVBAT1P addressed in this
paper. \vspace{2ex}

\section{Calibrating the deterministic barrier AT1P model }\label{sec:barrieroptions}
The fundamental hypothesis of the model we resume here is that the
underlying firm value process is a Geometric Brownian Motion
(GBM), which is also the kind of process commonly used for equity
stocks in the
Black Scholes model.

Classical structural models (Merton (1974), Black Cox (1976))
postulate a GBM (Black and Scholes) lognormal dynamics for the
value of the firm $V$. This lognormality assumption is considered
to be acceptable. Crouhy et al (2000) report that ``this
assumption is quite robust and, according to KMV's own empirical
studies, actual data conform quite well to this hypothesis.".

In these models the value of the firm $V$ is the sum of the firm
equity value $S$ and of the firm debt value $D$. The firm equity
value $S$, in particular, can be seen as a kind of (vanilla or
barrier-like) option on the value of the firm $V$. Merton
typically assumes a zero-coupon debt at a terminal maturity
${\bar{T}}$. Black Cox assume, besides a possible zero coupon
debt, safety covenants forcing the firm to declare bankruptcy and
pay back its debt with what is left as soon as the value of the
firm itself goes below a ``safety level" barrier. This is what
introduces the need for barrier option technology in structural
models for default.

In Brigo and Tarenghi (2004) the following proposition is proved:

\begin{proposition} {\bf (Analytically-Tractable
First Passage (AT1P) Model)} Assume the risk neutral dynamics for
the value of the firm $V$ is characterized by a risk free rate
$r(t)$, a payout ratio  $q(t)$ and an instantaneous volatility
$\sigma(t)$, according to
\[ dV(t) = V(t)\,(r(t)-q(t))\,dt+V(t)\,\sigma(t)\,dW(t)\] and assume a
safety barrier $\Hhat(t)$ of the form
\begin{equation}\label{Hhatbarrier} \Hhat(t)
=H\exp\left(-\int_0^t\left(q(s)-r(s) + (1+2\beta)
\frac{\sigma(s)^2}{2}\right)ds\right)\end{equation} where $\beta$
is a parameter that can be used to shape the safety barrier, $H$
is a reference level, and let $\tau$ be defined as the first time
where $V$ hits the safety covenants barrier $\Hhat$ from above,
starting from $V_0>H$,
\[ \tau = \inf\{ t \ge 0: V(t) \le \Hhat(t)\}. \]
Then the survival probability is given analytically by
\begin{eqnarray}\label{survcoher}
\mathbb{Q}\{\tau>T\} = \left[\Phi\left(\frac{\log \frac{V_0}{H}
+\beta\int_0^T \sigma(s)^2
ds}{\sqrt{\int_0^T\sigma(s)^2ds}}\right)-
\left(\frac{H}{V_0}\right)^{2\beta}\Phi\left(\frac{\log
\frac{H}{V_0} +\beta \int_0^T \sigma(s)^2
ds}{\sqrt{\int_0^T\sigma(s)^2ds}}\right)\right].
\end{eqnarray}
\end{proposition}

In presence of zero coupon debt for the maturity $\bar{T}$, the
actual default time is $\tau$ if $\tau < \bar{T}$, and is
$\bar{T}$ if $\tau \ge {\bar{T}}$ and the value of the firm at
$\bar{T}$ is below the debt face value. Otherwise, if $\tau \ge
{\bar{T}}$ and the value of the firm at $\bar{T}$ is above the
debt face value, there is no default.

In Brigo and Tarenghi (2004) we calibrate CDS quotes by means of
the above formula inserted in CDS valuation, backing out $t
\mapsto \sigma(t)$ from increasing-maturity CDS quotes through a
cascade inversion method. We now recall briefly the CDS payoff and
its risk neutral pricing formula.

One of the most representative protection instruments that can be
used against default is the Credit Default Swap (CDS). Consider
two companies ``A" (the {\em protection buyer}) and ``B" (the {\em
protection seller}) who agree on the following.

If a third reference company ``C" ({\em the reference credit})
defaults at a time $\tau_C \in (T_a,T_b]$, ``B" pays to ``A" at
time $\tau=\tau_C$ itself a certain ``protection" cash amount
$\lgd$ (Loss Given the Default of ``C"), supposed to be
deterministic in the present paper. This cash amount is a {\em
protection} for ``A" in case ``C" defaults. A typical stylized
case occurs when ``A" has bought a corporate bond issued from ``C"
and is waiting for the coupons and final notional payment from
this bond: If ``C" defaults before the corporate bond maturity,
``A" does not receive such payments. ``A" then goes to ``B" and
buys some protection against this risk, asking ``B" a payment that
roughly amounts to the bond notional in case ``C" defaults.
Typically $\lgd$ is equal to a notional amount, or to a notional
amount minus a recovery rate. We denote the recovery rate by
``$\rec$".

In exchange for this protection, company ``A" agrees to pay
periodically to ``B" a fixed ``running" amount $R$, at a set of
times $\{T_{a+1},\ldots,T_b\}$,  $\alpha_i = T_{i}-T_{i-1}$,
$T_0=0$. These payments constitute the ``premium leg" of the CDS
(as opposed to the $\lgd$ payment, which is termed the
``protection leg"), and $R$ is fixed in advance at time $0$; the
premium payments go on up to default time $\tau$ if this occurs
before maturity $T_b$, or until maturity $T_b$ if no default
occurs.
\[ \begin{array}{ccccc} \mbox{``B"} & \rightarrow & \mbox{ protection } \lgd \mbox{ at default $\tau_C$ if $T_a< \tau_C \le T_b$} & \rightarrow & \mbox{``A"} \\
%
\mbox{``B"} & \leftarrow  & \mbox{ rate } R \mbox{ at }
T_{a+1},\ldots,T_b \mbox{ or until default } \tau_C & \leftarrow &
\mbox{``A"}
\end{array} \]

Formally, we may write the RCDS (``R" stands for running)
discounted value at time $t$ seen from ``A" as
\begin{eqnarray}\label{discountedpayoffcds}
\nonumber\pcds_{a,b}(t) := - D(t,\tau) (\tau-T_{\beta(\tau)-1}) R
\mathbf{1}_{\{T_a < \tau < T_b \} } -\sum_{i=a+1}^b D(t,T_i)
\alpha_i R \mathbf{1}_{\{\tau \ge T_i\} }+\\ \mathbf{1}_{\{T_a <
\tau \le T_b \} }D(t,\tau) \lgd
\end{eqnarray}
where $t\in [T_{\beta(t)-1},T_{\beta(t)})$, i.e. $T_{\beta(t)}$ is
the first date among the $T_i$'s that follows $t$, and where
$\alpha_i$ is the year fraction between $T_{i-1}$ and $T_i$.
%
%
The pricing formula for this payoff depends on the assumptions on
the interest rates dynamics and on the default time $\tau$. Let
$\mathcal{F}_t$ denote the basic filtration without default,
typically representing the information flow of interest rates and
possibly other default-free market quantities (and also
intensities in the case of reduced form models), and
$\mathcal{G}_t = \mathcal{F}_t\vee \sigma\left(\{\tau<u\},u\leq
t\right)$ the extended filtration including explicit default
information. In our earlier ``structural model" AT1P framework
with deterministic default barrier the two sigma-algebras coincide
by construction, i.e. $\mathcal{G}_t = \mathcal{F}_t$, because
here the default is completely driven by default-free market
information. This is not the case with intensity models, where the
default is governed by an external random variable and
$\mathcal{F}_t$ is strictly included in $\mathcal{G}_t$, i.e.
$\mathcal{F}_t \subset \mathcal{G}_t$.

If we include barrier scenarios and take the barrier independent
of $W$, we are back in a situation where $\mathcal{F}_t \subset
\mathcal{G}_t$. Indeed, if for example we consider the random
barrier version with deterministic volatility and we know that
default occurs at a given time, say now, we also know that $V$ has
hit one of the barriers in the current scenario. By observing $V$,
we know which barrier scenario has realized itself, and by knowing
that default is now we know that the value of $V$ now is telling
us the value of $H$. But observation of $V$ alone (as happens with
${\cal F}$) until ``now" would not have allowed us to know $H$ now
or to know that default is ``now". Thus, contrary to the basic
AT1P model, ${\cal G}_t$ tells us more than ${\cal F}_t$ under
scenario based default barrier. The situation becomes more
involved if we allow for a random volatility $\sigma$, independent
of all observable quantities. We might assume the random
volatility is realized after a small time $\epsilon$, as we do in
smile modeling for example in Brigo, Mercurio and Rapisarda
(2004), but this leads to undesirable features such as unrealistic
future dynamics. While in smile modeling keeping $\sigma$ random
``forever" would have led us to an underlying equity or FX rate
that is not observed, unacceptably leading to payoffs that remain
random even at maturity, in case of structural models we may argue
that the current value of the firm is not necessarily known with
certainty. This would cut us some slack in keeping the volatility
in $V$ random ``forever". The mechanics of iterated expectation in
computing  prices does not change, one just has to make sure to
increase the inner conditioning filtration with the unobservable
variable and everything works fine. Here, however, even knowledge
of $\tau$ is not necessarily implying knowledge of $H$ or $V$,
given $V$'s uncertainty. At $\tau$ we would only know that $V$ hit
$H$, but since $V$ is now random, depending on the volatility
scenario, we cannot be sure of which $H$ scenario has been hit.
Thus the ${\cal G}$ filtration is less informative in this random
volatility case.

More generally, a discussion on the role of filtrations under
random default barriers and partial observation is in many papers
by Giesecke, see for example Giesecke and Goldberg (2004) (who
introduce the $I^2$ model).

We denote by $\cds_{a,b}(t, R, \lgd)$ the price at time $t$ of the
above standard running CDS. In general we can compute the CDS
price according to risk-neutral valuation (see for example
Bielecki and Rutkowski (2001)): \begin{equation}
\cds_{a,b}(t,R,\lgd) = \mathbb{E}\{\pcds_{a,b}(t)|\mathcal{G}_t\}
=:\mathbb{E}_t\{\pcds_{a,b}(t)\}\label{priceCDStheo}
\end{equation}
in our structural model setup. Straightforward computations lead
to the price at initial time $0$ as
\begin{eqnarray}\label{cdsformulaat1p}
\cds_{a,b}(0,R,\lgd) = R \int_{T_a}^{T_b} P(0,t)
(t-T_{\beta(t)-1}) d \Qx(\tau > t)\\ \nonumber - R \sum_{i=a+1}^b
P(0,T_i) \alpha_i \Qx(\tau \ge T_i)
 - \lgd \int_{T_a}^{T_b} P(0,t) d \Qx(\tau
> t)
\end{eqnarray}
so that if one has a formula for the curve of survival
probabilities $t \mapsto \Qx(\tau
> t)$, as in our AT1P structural model, one also has a formula for CDS.

A CDS is quoted through its ``fair" $R$, in that the rate $R$ that
is quoted by the market at time $t$
satisfies $\cds_{a,b}(t,R,\lgd) = 0$. 
The fair rate $R$ strongly depends on the default probabilities.
The idea is to use quoted values of these fair $R$'s with
different maturities to derive the default probabilities assessed
by the market.

While in simple intensity models the survival probabilities can be
interpreted as discount factors (with credit spreads as
discounting rates), and as such can be easily stripped from CDS's
or corporate bonds, in structural models the situation is much
more complicated.
%
%
In fact, here, it is not possible to find a simple ``credit
spread" formulation for $d\mathbb{Q}\{ \tau > t\}$ starting from
(\ref{survcoher}).

To calibrate the AT1P model to quoted market $R$'s for different
CDS, we insert the quoted $R$'s in Formula~(\ref{cdsformulaat1p})
and find the $\sigma$ and $H$ values that set said formula to
zero. See Brigo and Tarenghi (2004) for several numerical examples
and a case study on Parmalat CDS data.

\subsection{Numerical example}\label{sec:NumRes}
In this section we present some results of the calibration
performed with the AT1P structural model. We borrow from Brigo and
Tarenghi (2004). We consider CDS contracts having Vodafone as
underlying with recovery rate $\rec=40\%$ ($\lgd=0.6$). In
Table~\ref{VodSpread} we report the maturities $T_b$ of the
contracts and the corresponding ``mid" CDS rates $R_{0,b}^{\tiny
\mbox{MID}}(0)$ (quarterly paid) on the date of March 10th, 2004,
in basis points ($1bp = 10^{-4}$). We take $T_a=0$ in all cases.

\begin{table}[h!]
\begin{center}

\begin{tabular}{|c|c|c|c|c|}
\hline
           & CDS maturity $T_b$ & $R_{0,b}^{\tiny \mbox{BID}}(0)$ (bps) & $R_{0,b}^{\tiny \mbox{ASK}}(0)$ & $R_{0,b}^{\tiny \mbox{MID}}(0)$ \\
\hline
        1y &  20-mar-05 &         19 &         24 &       21.5 \\

        3y &  20-mar-07 &         32 &         34 &         33 \\

        5y &  20-mar-09 &         42 &         44 &         43 \\

        7y &  20-mar-11 &         45 &         53 &         49 \\

       10y &  20-mar-14 &         56 &         66 &         61 \\
\hline
\end{tabular}
\caption{\small Maturities of quoted CDS's with their
corresponding spreads on March 10, 2004.} \label{VodSpread}
\end{center}
\end{table}

\begin{table}[h!]
\begin{center}
\begin{tabular}{|c|cc|cc|}
\hline
   CDS mat & \multicolumn{ 2}{|c}{CDS value bid (bps)} & \multicolumn{ 2}{|c|}{CDS value ask (bps)} \\
\hline
        1y & \multicolumn{ 2}{|c}{2.56} & \multicolumn{ 2}{|c|}{-2.56} \\

        3y & \multicolumn{ 2}{|c}{2.93} & \multicolumn{ 2}{|c|}{-2.93} \\

        5y & \multicolumn{ 2}{|c}{4.67} & \multicolumn{ 2}{|c|}{-4.67} \\

        7y & \multicolumn{ 2}{|c}{24.94} & \multicolumn{ 2}{|c|}{-24.94} \\

       10y & \multicolumn{ 2}{|c}{41.14} & \multicolumn{ 2}{|c|}{-41.14} \\
\hline
\end{tabular}
\caption{\small CDS values computed with deterministic default
intensities stripped from mid $R$ quotes but with bid and ask
rates $R$ in the premium legs.} \label{VodSpreadbidask}
\end{center}
\end{table}

In Table~\ref{VodSpreadbidask} we report the values (in basis
points) of the CDS's computed inserting the bid and ask premium
rate $R$ quotes into the payoff and valuing the CDS with
deterministic intensities stripped by mid quotes. This way we
transfer the bid offer spread in the rates $R$ on a bid offer
spread on the CDS present value. In Table~\ref{VolaIntTab} we
present the results of the calibration performed with the
structural model and, as a comparison, of the calibration
performed with a deterministic intensity (credit spread) model
(using piecewise linear intensity). In this first example the
parameters used for the structural model have been selected on
qualitative considerations, and are $\beta = 0.5$ and $H/V_0=0.4$
(this is a significant choice since this value is in line with the
expected value of the random $H$, completely determined by market
quotes, in the Scenario based model presented later on).


We report the values of the calibrated parameters in the two
models (volatilities and intensities) and the survival
probabilities.

\begin{table}[h!]
\begin{center}
\begin{tabular}{|c||c|c||c|c|}
\hline $T_b$ & $\sigma$ & Surv. & Intensity & Surv. \\
\hline
 0   &   36.625\% &  100.000\%  &   0.357\% &  100.000\%\\

 1y  &   36.625\% &   99.627\%  &   0.357\% &   99.627\%\\

 3y  &   17.311\% &   98.316\%  &   0.952\% &   98.316\%\\

 5y  &   17.683\% &   96.355\%  &   1.033\% &   96.355\%\\

 7y  &   17.763\% &   94.206\%  &   1.189\% &   94.206\%\\

 10y &   21.861\% &   89.650\%  &   2.104\% &   89.604\%\\
 \hline
\end{tabular}
\end{center}
\caption{Results of the calibrations performed with both
models.}\label{VolaIntTab}
\end{table}

Comments on the realism of short term credit spreads and on the
robustness of default probabilities with respect to CDS are in
Brigo and Tarenghi (2004).


The parameter $\beta$ has not a direct economic meaning and in
principle can be chosen arbitrarily, and can help in setting more
or less stringent safety covenants, depending on one's attitude
towards default risk. As for $H$, in Brigo and Tarenghi (2004) we
present some methods to estimate it based on market
considerations.

\section{The scenario volatility/barrier SVBAT1P model}

In some cases it can be interesting to keep the volatility of the
process $V$ as an exogenous input coming from the equity and debt
worlds (for example it could be related to an historical or
implied volatility). Or we might retain a time-varying volatility
to be used only partly as a fitting parameter. Or, also, we might
wish to remove time-varying volatility to avoid an all-fitting
approach that is dangerous for robustness.

In all cases we would need to introduce other fitting parameters
into the model. One such possibility comes from introducing a
random default barrier. This corresponds to the intuition that the
balance sheet information is not certain, possibly because the
company is hiding some information. In this sense, assuming a
random default barrier can model this uncertainty. This means that
we retain the same model as before, but now the default barrier
level $H$ is replaced by a random variable $H$ assuming different
scenarios with given risk neutral probabilities. At a second
stage, we may introduce volatility scenarios as well. By imposing
time-constant volatility scenarios, in each single scenario we
loose flexibility with respect to AT1P, but regain flexibility
thanks to the multiple scenarios on otherwise too simple
time-constant volatilities. Even so, the scenario based model
results in a less flexible structure than the old deterministic
volatility and barrier AT1P model as far as CDS calibration is
concerned.

\newpage

In detail:

\begin{definition}{\bf (Scenario Volatility and Barrier Analytically Tractable 1st Passage model, SVBAT1P)}
Let the firm value process risk neutral dynamics be given by
\[ d V(t) = (r(t) - q(t)) V(t) dt + \nu(t) V(t) dW(t), \]
same notation as earlier in the paper. This time, however, let the
safety barrier parameter $H$ in~(\ref{Hhatbarrier}) and the firm
value volatility function $t \mapsto \nu(t)$ assume scenarios\\
$(H_1,t \mapsto \sigma^1(t)),\ldots,(H_{N-1},t\mapsto
\sigma^{N-1}(t)),(H_N,t\mapsto \sigma^N(t))$ with $\Qx$
probability $p_1,\ldots,p_{N-1},p_N$ respectively. The safety
barrier will thus be random and equal to
\[ \Hhat^i(t) := H_i \exp\left(-\int_0^t\left(q(s)-r(s) + (1+2\beta)
\frac{(\sigma^i(s))^2}{2}\right)ds\right)\] with probability
$p_i$. The random variables $H$ and $\nu$ are assumed to be
independent of the driving Brownian motion $W$ of the value of the
firm $V$ risk neutral dynamics.
\end{definition}

This definition has very general consequences. Indeed, if we are
to price a payoff $\Pi$ based on $V$, by iterated expectation we
have
\[ \Ex [ \Pi ] = \Ex\{ \Ex[ \Pi | H , \nu] \} = \sum_{i=1}^N p_i \Ex[ \Pi |
H=H_i, \nu = \sigma^i]\] Now, thanks to independence, the term
$\Ex[ \Pi | H=H_i, \nu = \sigma^i ]$ is simply the price of the
payoff $\Pi$ under the model with deterministic barrier and
volatility seen earlier in the paper, when the barrier parameter
$H$ is set to $H_i$ and the volatility to $t \mapsto \sigma^i(t)$,
so that the safety barrier is $\Hhat^i$. This means that, in
particular, for CDS payoffs we obtain
\begin{eqnarray} \cds_{a,b}(t,R,\lgd)  =  \sum_{i=1}^N
\cds_{a,b}(t,R,\lgd;H_i,\sigma^i)\cdot p_i\label{CDSvalue}
\end{eqnarray}
where $\cds_{a,b}(t,R,\lgd;H_i,\sigma^i)$ is the CDS price
(\ref{cdsformulaat1p}) computed according to survival
probabilities~(\ref{survcoher}) when the barrier $H$ is set to
$H_i$ and the volatility to $\sigma^i$. Let us consider now a set
of natural maturities for CDS quotes. This is to say that we
assume $T_a=0$ and $T_b$ ranging a set of standard maturities,
$T_b = 1y, 3y, 5y, 7y,10y$. Let us set
\[  \cds_{k}^i := \cds_{0,k}(0,R,\lgd; H_i,\sigma_i), \]
i.e. the CDS with first reset in $0$, final maturity $T_k$ and
valued under the deterministic barrier model with the barrier set
to $\Hhat^i$ and the deterministic volatility set to $t\mapsto
\sigma^i(t)$, which from now on we assume to be constant
$\sigma^i(t)=\sigma^i$ for all $t$.

Now assume we aim at calibrating the scenario based
barrier/volatility model to a term structure of CDS data. Our
first attempt is only with a scenario based barrier model, in that
we take a common, time constant volatility $\sigma^i =
\bar{\sigma}$ in all scenarios, so that the only uncertain
quantity in the model is the safety level barrier parameter $H$.

\subsection{CDS Calibration: Scenario based Barrier with Linear Algebra}

For this particular case we could reason with some linear algebra.
Assume we have $N$ CDS maturities $T_b$ to calibrate and consider
the parameters we might use. If we resort to a scenario based
barrier and fix the volatility exogenously, with $N$ barrier
scenarios we have $N$ values for $H$ and $N-1$ values for the
probabilities $p$, with a total of $2N-1$ parameters to fit $N$
quotes. Thus we have a system with more unknowns than equations.
One possible way of proceeding would be to fix $N-1$ parameters
(some barriers and some probabilities) and then looking for the
remaining free parameters. This procedure, when tried, is not very
stable and the calibration is quite difficult. A better way to
face the problem is the following. Let $C_{k,i} :=
\cds_{k}^i=\cds_{0,k}(0,R,\lgd; H_i,\bar{\sigma})$

If we write the relationship equating (\ref{CDSvalue}) in
correspondence of market $R$'s to zero, as should be, for $N$
maturities and with $N$ barriers, we have a linear system of the
form
\begin{equation}
{\bf C \cdot \underline{p} = \underline{0}}\label{lynearsystem}
\end{equation}
where ${\bf \underline{p}}$ is the ($N\times 1$) column vector  of
the probabilities and ${\bf \underline{0}}$ is a ($N\times 1$)
null vector. It is well known from basic algebra that if the
determinant of ${\bf C}$ is different from zero, then the system
(\ref{lynearsystem}) admits a unique solution and this solution is
identically equal to zero (the kernel of $C$ is the trivial null
space), which is not good as a set of probabilities ${\bf p}$.
Since we aim at a meaningful result for ${\bf p}$ we ask that
$\det({\bf C})=0$, so that the infinite solutions of the system
will be given by all the eigenvectors relative to the null
eigenvalue. Of all eigenvectors, we consider unitary norm
versions. Hence the first aim is to find those values of the
barriers that make the determinant of ${\bf C}$ equal to zero.
This is a single equation with $N$ unknowns. So we can fix $N-1$
barrier parameters $H$ and look for the last one solving this
``vanishing determinant" equation. Obviously we must be careful in
the choice of the parameters, since in principle given the first
$N-1$ barriers, the equation could admit no solutions. Also, even
if we find an admissible value for the last barrier, there is no
guarantee that the probabilities be all positive, since the
eigenvectors in principle could have negative components.

Actually, as one could guess beforehand, these considerations
result in a calibration technique that does not seem to be very
powerful. Let us analyze the particular case of the Vodafone
telecommunication company at the date of March 10th, 2004 (the
data are the mid quotes reported in Table \ref{VodSpread}; the
recovery rate is $\rec=40\%$, the common and time-constant
volatility in all scenarios is $\sigma=24\%$, $\beta$ is set equal
to 0.5 and the spreads are expressed in basis points). In the case
of deterministic barrier we find a term structure of deterministic
firm value volatility to calibrate exactly the data, as we have
seen in detail in Section~\ref{sec:NumRes}.

Here instead we keep the volatility fixed as an exogenous constant
and look for the barrier parameters $H$ and the probabilities $p$.

First of all we try to calibrate the first two maturities. We have
to fix the value for one of the barriers. We choose a high value,
i.e. $H_1=0.8$. The results of the calibration are reported in
Table \ref{2maturities}.

\begin{table}[h!]
{\small
\begin{center}
\begin{tabular}{|c|c|c|}
\hline
$i$ &   $H_i$   &   $p_i$   \\
\hline
1   & $0.3710^*$  &   $98.86\%$ \\
2   & $0.8000$    &   $1.14\%$  \\
\hline
\end{tabular}
\caption{Results of the calibration with the first two maturities.
The asterisk indicates that the corresponding value is the value
obtained from the calibration while the other barrier value has
been fixed in advance.} \label{2maturities}
\end{center}}
\end{table}

We see that the calibration assigns a low probability to the high
barrier and a high probability to the low barrier: This is
consistent with the fact that the CDS rates are very low,
indicating a quite good credit quality of the underlying. In fact
the high barrier (corresponding to high debt) indicates proximity
to default, while the low barrier (low debt) indicates a distant
default possibility.

Now let us add a maturity, i.e. let us move from two maturities to
three maturities, namely the first three dates in Table
\ref{VodSpread}. It would be somehow natural to try a
configuration of barriers which is an extension of the previous
one, i.e. to fix the two barriers just found and to look for a
third one. However, this procedure does not lead to a solution: To
 calibrate we need to change one of the barriers. We decided to keep
fixed $H=0.8$ and change the other one, choosing $H=0.2$. The
results are reported in Table \ref{3maturities}.

\begin{table}[h!]
{\small
\begin{center}
\begin{tabular}{|c|c|c|}
\hline
$i$ &   $H_i$   &   $p_i$   \\
\hline
1   & $0.2000$    &   $61.06\%$ \\
2   & $0.4303^*$  &   $37.83\%$ \\
3   & $0.8000$    &   $1.10\%$  \\
\hline
\end{tabular}
\caption{Results of the calibration with the first three
maturities. The asterisk indicates that the corresponding value is
the value obtained from the calibration while the other values
have been fixed in advance.} \label{3maturities}
\end{center}}
\end{table}

We notice that a splitting has occurred: From a single barrier set
at 0.3710 we obtained two barriers, one higher and one lower. The
highest barrier has still a low probability, almost equal to the
corresponding one obtained earlier, when calibrating only two
maturities. The other probability of the two-maturities case has
split, following the splitting of the $H$ parameters, and now the
lowest $H$ is the most likely.

Again, moving to the first four maturities in Table
\ref{VodSpread} we would try simply to add a new barrier parameter
while retaining the $H$'s calibrated under three quotes. As
before, this calibration does not work. Then we substitute the
last barrier found (0.4303) with 0.5500 before trying the
calibration. In Table \ref{4maturities} we report the result of
the calibration.

\begin{table}[h!]
{\small
\begin{center}
\begin{tabular}{|c|c|c|}
\hline
$i$ &   $H_i$   &   $p_i$   \\
\hline
1   & $0.2000$    &   $27.65\%$ \\
2   & $0.3347^*$  &   $66.75\%$ \\
3   & $0.5500$    &   $7.71\%$  \\
3   & $0.8000$    &   $0.89\%$  \\
\hline
\end{tabular}
\caption{Results of the calibration with the first four
maturities. The asterisk indicates that the corresponding value is
the value obtained from the calibration while the other values
have been fixed in advance.} \label{4maturities}
\end{center}}
\end{table}

As when moving from two to three maturities, when moving from
three to four maturities we observe a splitting of one barrier
parameter $H$ into two values, one higher and one lower.  We see
that the greatest part of the probability is divided among the
three ``low" barriers while the highest barrier has still a low
probability. More precisely, we see that adding information (i.e.
adding maturities), the high barrier seems to become less
probable. In general adding information leads obviously to a
refinement of the barrier configuration.

It may be interesting to see what happens in case of a worse
credit quality. To see this let us arbitrarily change the CDS
rates (all other things being equal), in particular let us double
them, as shown in Table \ref{VodSpreadBis}, and repeat the
calibration as before. The results are reported in Tables
\ref{2maturitiesBis} and \ref{3maturitiesBis}.

\begin{table}[h!]
{\small
\begin{center}
\begin{tabular}{|c|c|}
\hline
Maturity $T_b$ & Doubled Rate $R_{0,b}^{\tiny \mbox{MID}}(0)$ (bps)\\
\hline
March 21st, 2005 & 43.0 \\
March 20th, 2007 & 66.0 \\
March 20th, 2009 & 86.0 \\
\hline
\end{tabular}
\caption{Maturities of quoted CDS's with their corresponding
doubled spreads, valuation date March 10, 2004.}
\label{VodSpreadBis}
\end{center}}
\end{table}

\begin{table}[h!]
{\small
\begin{center}
\begin{tabular}{|c|c|c|}
\hline
$i$ &   $H_i$   &   $p_i$   \\
\hline
1   & $0.4123^*$  &   $97.76\%$ \\
2   & $0.8000$    &   $2.24\%$  \\
\hline
\end{tabular}
\caption{Results of the calibration with the first two maturities.
The asterisk indicates that the corresponding value is the value
obtained from the calibration while the other barrier value has
been fixed in advance.} \label{2maturitiesBis}
\end{center}}
\end{table}

\begin{table}[h!]
{\small
\begin{center}
\begin{tabular}{|c|c|c|}
\hline
$i$ &   $H_i$   &   $p_i$   \\
\hline
1   & $0.2000$    &   $19.81\%$ \\
2   & $0.4277^*$  &   $77.98\%$ \\
3   & $0.8000$    &   $2.21\%$  \\
\hline
\end{tabular}
\caption{Results of the calibration with the first three
maturities. The asterisk indicates that the corresponding value is
the value obtained from the calibration while the other values
have been fixed in advance.} \label{3maturitiesBis}
\end{center}}
\end{table}

We see that by increasing the CDS rates, i.e. decreasing the
credit quality of the underlying firm, the calibrated barriers
parameters $H$ become higher and also the probabilities to have
high barriers increase, as one would expect, since high barriers
increase the proximity of default.

With this method we have not been able, by adding a further
scenario to $H$, to calibrate all of the five CDS of
Table~\ref{VodSpread}. This is due probably to the
time-homogeneity of the model. To have a larger calibration
capability we may have to use a time varying~$\nu$.

To sum up, as one could have expected, this ``linear algebra"
procedure and the scenario based barrier approach in general are
not very robust and are quite sensitive to CDS rates: For high
rates the calibration works only for a low number of maturities,
and the refinement is limited. This is one of the main drawbacks
of the model which makes it less powerful than the standard
calibration with deterministic barrier AT1P of Brigo and Tarenghi
(2004), at least if one insists in having an exact calibration.
One possibility to improve the model is to include a time
dependency for the parameters (for example in the volatility
$\bar{\sigma}$), as hinted at before. This idea is currently under
investigation. A different possibility is moving in two different
directions: first, abandoning the ``linear algebra" approach and
using a numerical optimization within the same framework, and
secondly abandoning the single time-constant volatility scenario
and allowing several time-constant volatility scenarios to come
into play, which we explore in Section~\ref{barrvolscen}.

\subsection{CDS Calibration: Scenario based Barrier with Numerical Optimization}
Here we abandon the requirement to obtain a perfect calibration
and allow for a (small) calibration error, while trying to restore
a one to one correspondence between model parameters and market
quotes.

We start by the first three quotes in Table~\ref{VodSpread}, but
this time calibrate them with an optimization, abandoning the
``linear algebra" approach. We minimize the function expressing
the sum of the squares of the CDS's prices in the model, since
each CDS price should be zero in correspondence of the market
quoted $R$. Thus we actually solve numerically
\begin{eqnarray*}
[H_1^\ast,H_2^\ast,p_1^\ast] & = & \mbox{argmin}_{H,p}
\sum_{k=1}^3 [ p_1 \cds_{0,k}(0,R^{\mbox{\tiny
MID}}_{0,k}(0),\lgd;H_1)\\ &&+(1-p_1) \cds_{0,k}(0,R^{\mbox{\tiny
MID}}_{0,k}(0),\lgd;H_2)]^2
\end{eqnarray*}
We find the following results:

\begin{table}[h!]
\begin{center}
\begin{tabular}{|c|c|}
\hline
   $H_i$ & $p_i$ \\
\hline
    0.3188 &    94.83\% \\

    0.6592 &     5.17\% \\
\hline
\end{tabular}\caption{SBAT1P calibrated parameters. Exact
calibration of the first three CDS quotes of
Table~\ref{VodSpread}. The expected value of the barrier is
$\mathbb{E}[H]=0.3364$.}\label{sbat1p3cdscal}
\end{center}
\end{table}

Luckily, in this case the target function is practically zero, so
that the calibration is exact. While this could be somehow hoped,
since the number of free parameters matches the number of quotes
to be calibrated, in general it is not guaranteed. Indeed, we will
see shortly that even if the parameters are in the right number
they often do not have enough flexibility to account for the
market quotes exactly.

Let us move to optimizing over the five quotes of
Table~\ref{VodSpread}. To this end, we can choose the version with
three $H$ scenarios and two probabilities. We run an optimization
to solve
\begin{eqnarray*}
[H_1^\ast,H_2^\ast,H_3^\ast,p_1^\ast,p_2^\ast] & = &
\mbox{argmin}_{H,p} \sum_{k=1}^5 \bigg{[} p_1
\cds_{0,k}(0,R^{\mbox{\tiny MID}}_{0,k}(0),\lgd;H_1)\\ & & + p_2
\cds_{0,k}(0,R^{\mbox{\tiny MID}}_{0,k}(0),\lgd;H_2)  +
p_3\cds_{0,k}(0,R^{\mbox{\tiny MID}}_{0,k}(0),\lgd;H_3)\bigg{]}^2
\end{eqnarray*}
where in the optimization we impose the $p$ to take values that
are allowed for probabilities, i.e. in $[0,1]$ and adding up to
one. This time we obtain

\begin{table}[!h]
\begin{center}
\begin{tabular}{|c|c|}
\hline
   $H_i$ & $p_i$ \\
\hline
    0.7296 &     1.24\% \\

    0.3384 &    97.52\% \\

    0.7296 &     1.24\% \\
\hline
\end{tabular}\caption{SBAT1P calibrated parameters. Calibration of the first five CDS quotes of
Table~\ref{VodSpread}. The expected value of the barrier is
$\mathbb{E}[H]=0.3481$.}\label{sbat1p5cdscal}
\end{center}
\end{table}

The calibration error is not zero this time, but amounts to $915
bps^2$. We will discuss the calibration error shortly. One barrier
scenario $H$ remains close to 0.3188, with an even higher
probability than before. Notice an important point: the two other
scenarios of $H$ more in proximity of default are identical, being
both 0.7296, with equal probabilities 0.0124. This suggests the
parametrization to be not effective in describing a dynamics
consistent with the cross sectional data we are observing, since
the parameters collapse to a sub-parameterization with just two
barrier scenarios (three parameters in total) when fitting five
quotes. This seems to suggest the new parameters we added with
respect to the previous case add nothing to explain the increase
of data. And indeed, if one tries the model with two barriers and
one probability to fit the five quotes above, consistently with
what we just obtained one finds the same calibration error and the
parameters

\begin{table}[!h]
\begin{center}
\begin{tabular}{|c|c|}
\hline
   barrier & probability \\
\hline
    0.7296 &     2.48\% \\

    0.3384 &    97.52\% \\
\hline
\end{tabular}
\caption{SBAT1P calibrated parameters. Calibration of the first
five CDS quotes of Table~\ref{VodSpread} with only three unknowns
parameters.}\label{sbat1p5cdscalbis}
\end{center}
\end{table}

Let us have a look at the calibration error in single CDS's. If we
compute the five CDS prices using the $R$ mid quotes and with the
parameters $H$ and $p$ above coming out of the optimization, we
obtain the values reported in Table~\ref{CDSvaluesSbat}.

\begin{table}[!h]
\begin{center}
\begin{tabular}{|c|cc|}
\hline
   CDS maturity $T_k$ & \multicolumn{ 2}{|c|}{$\cds_{0,k}(0,R_{0,k}^{\mbox{\tiny MID}},H_1,H_2,H_3;p_1,p_2,p_3)$ (bps)} \\
\hline
        1y & \multicolumn{ 2}{|c|}{-2.77} \\

        3y & \multicolumn{ 2}{|c|}{9.99} \\

        5y & \multicolumn{ 2}{|c|}{-1.47} \\

        7y & \multicolumn{ 2}{|c|}{-22.99} \\

       10y & \multicolumn{ 2}{|c|}{16.63} \\
\hline
\end{tabular}
\end{center}\caption{CDS values obtained using the parameters
resulting from the calibration.}\label{CDSvaluesSbat}
\end{table}

We can compare these calibration errors on the single CDS's
present values with the errors induced in the CDS present value by
the market $R$ bid-ask spreads, reported in
Table~\ref{VodSpreadbidask}. The first two CDS are out of the bid
ask windows with our calibration, the worst case being the second,
giving $9.99$ bps of present value against a corresponding bid ask
window $[-2.93, 2.93]$.

This situation cannot be considered to be fully satisfactory,
which leads us to the following extension.

\subsection{Scenario based Barrier and Volatility: Numerical
Optimization}\label{barrvolscen}

Now we consider again all of the five quotes of
Table~\ref{VodSpread} and we use the general model with two
scenarios on the barrier/volatility parameters  $(H,\nu)$ and one
probability (the other one being determined by normalization to
one), with a total of five parameters for five quotes. Here, by
trial and error we decided to set $\beta=0$.
\begin{eqnarray*}
[H_1^\ast,H_2^\ast; \sigma^1_\ast,\sigma^2_\ast;p_1^\ast] &=&
\mbox{argmin}_{H,p,\nu} \sum_{k=1}^5 \bigg{[} p_1
\cds_{0,k}(0,R^{\mbox{\tiny MID}}_{0,k}(0),\lgd;H_1,\sigma^1)\\
& & +(1-p_1) \cds_{0,k}(0,R^{\mbox{\tiny
MID}}_{0,k}(0),\lgd;H_2,\sigma^2)\bigg{]}^2
\end{eqnarray*}

\begin{table}[h!]
\begin{center}
\begin{tabular}{|c|c|c|}
\hline
        $H_i$ &         $\sigma_i$ &         $p_i$ \\
\hline
    0.3721 &    17.37\% &    93.87\% \\

    0.6353 &    23.34\% &     6.13\% \\
\hline
\end{tabular}\caption{SVBAT1P model calibrated to Vodafone CDS. The expected value of the barrier is
$\mathbb{E}[H]=0.3882$.}\label{tablesvbat1pvodafone}
\end{center}
\end{table}

We obtain a much lower optimization error than before, i.e. $147
bps^2$, corresponding to the following calibration errors on
single CDS present values:

\begin{table}[!h]
\begin{center}
\begin{tabular}{|c|cc|}
\hline
CDS maturity $T_k$ & \multicolumn{ 2}{|c|}{$\cds_{0,k}(0,R_{0,k}^{\mbox{\tiny MID}},H_1,H_2;\sigma^1,\sigma^2;p_1)$ (bps)} \\
\hline
1y & \multicolumn{ 2}{|c|}{1.38} \\
3y & \multicolumn{ 2}{|c|}{-3.89} \\
5y & \multicolumn{ 2}{|c|}{8.16} \\
7y & \multicolumn{ 2}{|c|}{-7.56} \\
10y & \multicolumn{ 2}{|c|}{2.41} \\
\hline
\end{tabular}
\end{center}\caption{CDS values obtained using the parameters resulting from
the calibration.}\label{CDSvaluesSVbat}
\end{table}

Now the single CDS calibration errors are much lower than before,
being the CDS present values corresponding to market $R$ closer to
zero, with the exception of the five years maturity, which could
be adjusted by introducing weights in the target function.

In particular we introduce weights that are inversely proportional
to the bid-ask spread and repeat the calibration for the SVBAT1P
model finding the results presented in
Tables~\ref{tablesvbat1pvodafoneweighted}
and~\ref{CDSvaluesSVbatweighted} (compare with
Tables~\ref{tablesvbat1pvodafone} and~\ref{CDSvaluesSVbat}).

\begin{table}[h!]
\begin{center}
\begin{tabular}{|c|c|c|}
\hline
$H_i$ &         $\sigma_i$ &         $p_i$ \\
\hline
0.3713 &    17.22\% &    92.63\% \\
0.6239 &    22.17\% &     7.37\% \\
\hline
\end{tabular}\caption{SVBAT1P model calibrated to Vodafone CDS using weights in the objective
function which are inversely proportional to the bid ask-spread.
The expected value of the barrier is
$\mathbb{E}[H]=0.3899$.}\label{tablesvbat1pvodafoneweighted}
\end{center}
\end{table}

\begin{table}[!h]
\begin{center}
\begin{tabular}{|c|cc|}
\hline
CDS maturity $T_k$ & \multicolumn{ 2}{|c|}{$\cds_{0,k}(0,R_{0,k}^{\mbox{\tiny MID}},H_1,H_2;\sigma^1,\sigma^2;p_1)$ (bps)} \\
\hline
1y & \multicolumn{ 2}{|c|}{5.85} \\
3y & \multicolumn{ 2}{|c|}{-3.76} \\
5y & \multicolumn{ 2}{|c|}{4.92} \\
7y & \multicolumn{ 2}{|c|}{-10.46} \\
10y & \multicolumn{ 2}{|c|}{1.47} \\
\hline
\end{tabular}
\end{center}\caption{CDS values obtained using the parameters resulting from
the weighted calibration.}\label{CDSvaluesSVbatweighted}
\end{table}

We see that now the 1y CDS value is outside the bid-ask spread,
but conversely the 5y CDS value has decreased, assuming a size
more in line with the spread. This is what we aimed at, since the
5y CDS is probably the most liquid one and the model needs to
reproduce it well.

\begin{figure}[!h]
\begin{center}
\includegraphics[angle=270,width=0.7\textwidth]{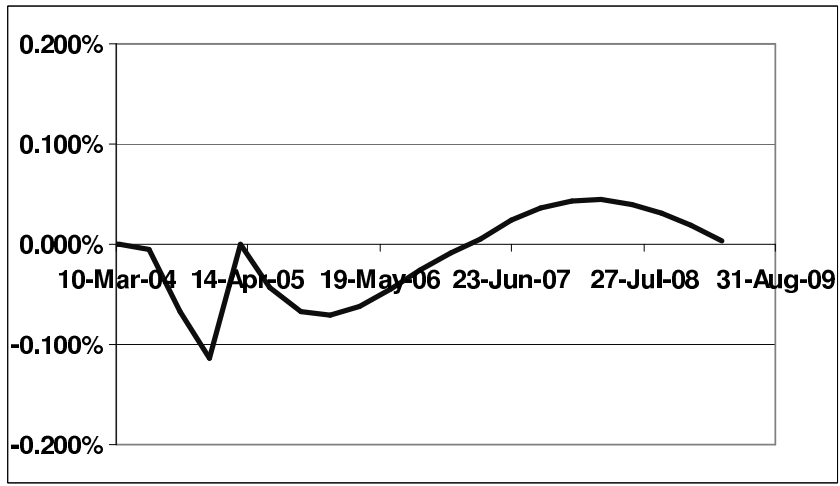}
\end{center}
\caption{\small SBAT1P-AT1P survival probabilities difference. The
data for the SBAT1P are those in Table~\ref{sbat1p3cdscal} and
refer to an exact calibration of 3 CDS
quotes.}\label{SBATprobExact}
\end{figure}

\begin{figure}[!h]
\begin{center}
\includegraphics[angle=270,width=0.7\textwidth]{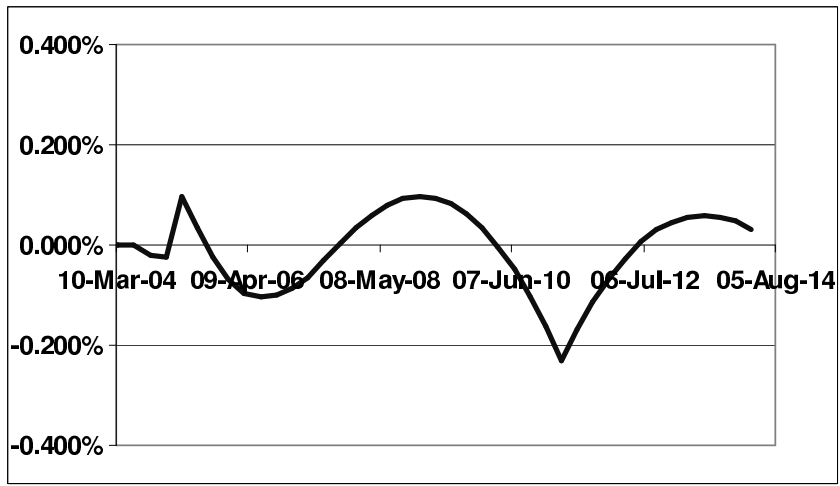}
\end{center}
\caption{\small SVBAT1P-AT1P survival probabilities difference.
The data for the SVBAT1P are those in
Table~\ref{tablesvbat1pvodafoneweighted} and refer to an optimal
minimum calibration of 5 CDS quotes using weights in the objective
function.}\label{SVBATprobMin}
\end{figure}

Furthermore, to see the goodness of the calibration, we can
compare the survival probabilities of the S(V)BAT1P model with
those of the perfectly calibrated AT1P model. In
Figure~\ref{SBATprobExact} we plot the difference between the
survival probabilities computed with the SBAT1P model exact
calibration (see Table~\ref{sbat1p3cdscal}) and the AT1P
calibration (on a 5 years horizon). In Figure~\ref{SVBATprobMin}
we plot the difference between the survival probabilities computed
with the SVBAT1P model weighted calibration (see
Table~\ref{tablesvbat1pvodafoneweighted}) and the AT1P calibration
(on a 10 years horizon). The two figures show a good agreement
between the models in terms of probabilities, and the main
difference is that the survival probabilities nearly coincide at
CDS maturities in the case of exact calibration
(Figure~\ref{SBATprobExact}), while a difference is found in the
minimization case (Figure~\ref{SVBATprobMin}). Anyway, it is clear
that the fit is quite good, indicating two main results: (i) the
SVBAT1P model returns a good fit of CDS quotes, even if the
calibration is not perfect; (ii) as already observed in Brigo and
Tarenghi (2004), CDS are a robust source for survival (default)
probabilities, whatever model is used.

More in general, when combining default barrier and volatility
scenarios, scenarios on $\nu$ and $H$ can be taken jointly (as we
did so far) or separately, but one needs to keep the combinatorial
explosion under control. The pricing formula remains easy, giving
linear combination of formulas in each basic scenario.   Taking
time-varying parametric forms for the $\sigma^i$'s can add
flexibility and increase the calibrating power of the model, and
will be addressed in future work.

A final important remark is in order. We see that the parameters
resulting from the scenario versions calibration are more credible
than those obtained in the AT1P framework. In particular we notice
that in all cases we have an expected $H$ which is comparable with
the fixed $H=0.4$ used in the deterministic case. What is more,
even if the $H$ are similar, the volatilities involved are quite
smaller than before. As previously hinted at, this fact is
essentially due to volatility scenarios inducing a mixture of
lognormal distributions for the firm value, implying fatter tails,
and allowing for the same default probabilities with smaller
volatilities.

\section{Counterparty risk in equity swap pricing: Reprise}

{\em ``Of course it does. \\A conservative estimate puts
material reality as ten thousand million light years across.\\
Even using the combined computational function of all the
processors in the united planets capable of dealing with numerical
scale, it would
take at least ninety years to..."}\\

Brainiac 5, ``Legion Lost", 2000, DC Comics\\

This section summarizes the results on counterparty risk pricing
in Equity Return Swaps under AT1P in Brigo and Tarenghi (2004) and
presents a short analysis based on the SBVAT1P models introduced
in this paper. This is an example of counterparty risk pricing
with the calibrated structural model.

Let us consider an equity return swap payoff. Assume we are a
company ``A" entering a contract with company ``B", our
counterparty. The reference underlying equity is company ``C". The
contract, in its prototypical form, is built as follows. Companies
``A" and ``B" agree on a certain amount $K$ of stocks of a
reference entity ``C" (with price $S=S^C$) to be taken as nominal
($N=K\,S_0$). The contract starts in $T_a=0$ and has final
maturity $T_b = T$. At $t=0$ there is no exchange of cash
(alternatively, we can think that ``B" delivers to ``A" an amount
$K$ of ``C" stock and receives a cash amount equal to $K S_0$). At
intermediate times ``A" pays to ``B" the dividend flows of the
stocks (if any) in exchange for a periodic rate (for example a
semi-annual LIBOR or EURIBOR rate $L$) plus a spread $X$. At final
maturity $T=T_b$, ``A" pays $K S_T$ to ``B" (or gives back the
amount $K$ of stocks) and receives a payment $K S_0$. This can be
summarized as follows:

\begin{center}
Initial Time 0: no flows, or\\  A $\longrightarrow$ $K S^C_0$ cash
$\longrightarrow$ B\\
 A $ \longleftarrow$  $K$ equity of ``C" $\longleftarrow$ B\\
....  \\
Time $T_i$:\\ A $\longrightarrow$ equity dividends of ``C" $\longrightarrow$ B\\
A $\longleftarrow$  Libor + Spread $\longleftarrow$ B\\
.... \\
Final Time $T_b$:\\  A $\longrightarrow$ K equity of ``C"
$\longrightarrow$ B
\\
A $\longleftarrow$  $K S^C_0$ cash $\longleftarrow$ B \end{center}

The price of this product can be derived using risk neutral
valuation, and the (fair) spread is chosen in order to obtain a
contract whose value at inception is zero. We ignore default of
the underlying ``C", thus assuming it has a much stronger credit
quality than the counterparty ``B". It can be proved that if we do
not consider default risk for ``B" either, the fair spread is
identically equal to zero. But when taking into account
counterparty default risk in the valuation the fair spread is no
longer zero.   In case an early default of the counterparty ``B"
occurs, the following happens.  Let us call $\tau=\tau_B$ the
default instant. Before $\tau$ everything is as before,  but if
$\tau\leq T$, the net present value (NPV) of the position at time
$\tau$ is computed. If this NPV is negative for us, i.e. for ``A",
then its opposite is completely paid to ``B" by us at time $\tau$
itself. On the contrary, if it is positive for ``A", it is not
received completely but only a recovery fraction $\rec$ of that
NPV is received by us. It is clear that to us (``A") the
counterparty risk is a problem when the NPV is large and positive,
since in case ``B" defaults we receive only a fraction of it.

The risk neutral expectation of the discounted payoff is given in
the following proposition (see e.g. Brigo and Tarenghi (2004),
$L(S,T)$ is the simply compounded rate at time $S$ for maturity
$T$):
\begin{proposition}{\bf (Equity Return Swap price under
Counterparty Risk)}. The fair price of the Equity Return Swap
defined above can be simplified as follows:
\begin{eqnarray*}
\mbox{ERS}(0) = KS_0 X \sum_{i=1}^b\alpha_i
P(0,T_i)-\lgd\,\mathbb{E}_0\bigg\{ \mathbf{1}_{\{\tau\leq
T_b\}}D(0,\tau)(\npv(\tau))^+\bigg\}.
\end{eqnarray*}
where
\begin{eqnarray}\label{NPVdef}
\npv(\tau) & = &
\mathbb{E}_{\tau}\bigg\{-K\,\npv_{dividends}^{\tau\div
T_b}(\tau)+KS_0\,\sum_{i=\beta(\tau)}^{b}
D(\tau,T_i)\alpha_i\left(L(T_{i-1},T_i)+X\right) \nonumber \\
& & +\left(KS_0-KS_{T_b}\right)D\left(\tau,T_b\right)\bigg\}.
\end{eqnarray}
and where we denote by $\npv_{dividends}^{s \div t}(u)$  the net
present value of the dividend flows between $s$ and $t$ computed
in $u$.

The first term in $\Pi_{ES}$ is the equity swap price in a
default-free world, whereas the second one is the optional price
component due to counterparty risk.
\end{proposition}

If we try and find the above price by computing the expectation
through a Monte Carlo simulation, we have to simulate both the
behavior of $S_t$ for the equity ``C" underlying the swap, and the
default of the counterparty ``B". In particular we need to know
exactly $\tau=\tau_B$. Obviously the correlation between ``B" and
``C" could have a relevant impact on the contract value. Here the
structural model can be helpful: Suppose to calibrate the
underlying process $V$ to CDS's for name ``B", finding the
appropriate default barrier and volatilities according to the
procedure outlined earlier in this paper with the AT1P model. We
could set a correlation between the processes $V^B_t$ for ``B" and
$S_t$ for ``C", derived for example through historical estimation
directly based on equity returns, and simulate the joint evolution
of $[V^B_t, S_t]$. As a proxy of the correlation between these two
quantities we may consider the correlation between $S^B_t$ and
$S^C_t$, i.e. between equities.

Going back to our equity swap, now it is possible to run the Monte
Carlo simulation, looking for the spread $X$ that makes the
contract fair.

We performed some simulations under different assumptions on the
correlation between ``B" and ``C". We considered five cases:
$\rho= -1$, $\rho = -0.2$, $\rho = 0$, $\rho = 0.5$ and $\rho =
1$. In Table~\ref{EScorrelation} we present the results of the
simulation, together with the error given by one standard
deviation (Monte Carlo standard error). For counterparty ``B" we
used the Vodafone CDS rates seen earlier. For the reference stock
``C" we used a hypothetical stock with initial price $S_0=20$,
volatility $\sigma = 20\%$ and constant dividend yield $q=0.80\%$.
The contract has maturity $T=5y$ and the settlement of the LIBOR
rate has a semi-annual frequency. Finally, we included a recovery
rate $\rec=40\%$. The starting date is the same we used for the
calibration, i.e. March 10th, 2004. Since the reference number of
stocks $K$ is just a constant multiplying the whole payoff,
without losing generality we set it equal to one.

In order to reduce the errors of the simulations, we have adopted
a variance reduction technique using the default indicator (whose
expected value is the known default probability) as a control
variate. In particular we have used the default indicator
$1_{\{\tau \leq T\}}$ at the maturity $T$ of the contract, which
has a large correlation with the final payoff. Even so, a large
number of scenarios is needed to obtain errors with a lower order
of magnitude than $X$. In our simulations we have used $N =
2000000$.

We notice that $X$ increases together with $\rho$, and in Brigo
and Tarenghi (2004) we explain why this is natural.

\begin{table}[h!]
\begin{center}
\begin{tabular}{|c|c|c|c|}
\hline $\rho$ & X & ES payoff & MC error \\
\hline
-1      &   0       &   0       &   $0$   \\
-0.2    &   2.45    &   -0.02   &   1.71    \\
0       &   4.87    &   -0.90   &   2.32    \\
0.5     &   14.2    &   -0.53   &   2.71    \\
1       &   24.4    &   -0.34   &   0.72    \\
\hline
\end{tabular}
\end{center}
\caption{\small Spread $X$ (in bps) under five correlations,
$S_0=20$, basic AT1P model. We also report the value of the
average of the simulated payoff (times 10000) across the $2000000$
scenarios and its standard error, thus showing that $X$ is fair
(leads to a zero NPV).}\label{EScorrelation}
\end{table}

To check the impact of the scenarios barrier, we have re-priced
with the same $X$'s found in Table~\ref{EScorrelation} for AT1P
our equity swap under the SBAT1P model calibrated to the same CDS
data, i.e. with the parameters given in Table~\ref{sbat1p3cdscal}.
We have the outputs given in Table~\ref{essvbat1p}. If the models
are close in terms of equity swap pricing, this payoff should be
zero. We find actually values ranging from $-28$bps to $165$ bps.
Recalling that $S_0 = 20$, for a unit notional we would get a
maximum discrepancy of about $165/20\approx 8$ bps. By comparing
with the bid-offer window induced by market bid-offer $R$'s on
CDS's with unit notional, as shown for example in
Table~\ref{VodSpreadbidask}, we realize that this difference, if
not completely negligible, is small. Besides, for the largest
payoff value, i.e. $\rho=0.5$ (where the AT1P model gives -0.53
bps and the SBAT1P gives 165.5 bps)  we notice that with $X=16$
(instead of 14.2) the AT1P model would give a payoff price of
166.9, so that our price difference between SBAT1P and AT1P, when
translated back in AT1P spread $X$, is less than two basis points.

\begin{table}[h!]
\begin{center}
\begin{tabular}{|c|c|c|c|}
\hline $\rho$ & X & ES payoff & MC error estimate\\
\hline
-0.2    &   2.45    &   -28.44  &   1.49    \\
0       &   4.87    &   3.45    &   2.04    \\
0.5     &   14.2    &   165.50  &   2.29    \\
\hline
\end{tabular}\caption{Equity swap valuation under the SBAT1P
model calibrated to the same CDS data as the AT1P model leading to
Table~\ref{EScorrelation}. }\label{essvbat1p}
\end{center}
\end{table}

We can try the same considerations with the SVBAT1P model, with
the weighted calibration given in
Table~\ref{tablesvbat1pvodafoneweighted}. If we consider the case
with $\rho=0.5$, the Monte Carlo method gives us the payoff
expected value as 292.03 bps, with a Monte Carlo error of about
1.67 bps. Again, recalling that $S_0 = 20$ we can consider
$292.03/20 \approx 14.6$, and compare with CDS payoff bid ask
values, as in Table~\ref{VodSpreadbidask}. We see that we are
within the 7y and 10y CDS bid ask spreads but not within the 1y,
3y and 5y spreads. Also in this case we tried to see which value
for $X$ in the AT1P model with $\rho=0.5$ would produce an
expected payoff close to 290. We obtained that a spread of
$X=17.3$ would give an expected payoff value of 289, so that we
see that the difference between AT1P and SVBAT1P in terms of AT1P
spread is about 17.3-14.2= 3.1 bps. By comparing with bid-ask
spreads in CDS rates $R$, as given in Table~\ref{VodSpread}, we
see again that this difference is inside the 1y, 7y and 10y
spreads on the $R$ quotes (which are respectively of 5, 8 an 10
bps). Also, even if a little larger, it is comparable to the 3y
and 5y spreads (both of 2 bps), which are very narrow being
related to the most traded CDS maturities. Thus we see that the
difference is nearly negligible.

In general we have deviated further, since we started from the 0
payoff expected value under AT1P and the 165.5 expected payoff
value under SBAT1P. Since AT1P, SBAT1P and SVBAT1P are calibrated
to the same CDS data up to five years (but SVBAT1P is also
calibrated to 7y and 10y CDS), we are seeing here that the
different dynamics assumptions in the three models lead to
different counterparty risk valuations in the equity return swap.
The difference is not large when compared to bid ask spreads of
CDS. We may expect more significant deviations in hedging. We have
to keep in mind an important consideration, though. SVBAT1P is
calibrated not exactly and not only on 1y, 3y and 5y CDS as the
earlier AT1P and SBAT1P models. Probably some of the difference
between the price obtained with the SBAT1P and the SVBAT1P is to
be attributed to this fact.

\begin{remark}{\bf (Nested calibration with AT1P but not with S(V)BAT1P).} A further important remark is the
``nested calibration" aspect. With AT1P the calibration is
``nested", in that adding one CDS with a larger maturity does not
change the earlier $\sigma$ parameters found with the calibration
up to that point. In a way, this is a ``cascade calibration". This
may be helpful with sensitivities and bucketing. Instead, in the
S(V)BAT1P model the parameters assume a ``global" role: if we add
a CDS quote and recalibrate, all the parameters change again. In
principle, every CDS quote has an impact on {\em all} the
S(V)BAT1P model parameters at the same time. This is less
desirable in computing sensitivities to market inputs and can lead
to numerical problems, especially with SVBAT1P where a further
problem is that the calibration is not exact.
\end{remark}

We will address these matters in future work.

\section{Conclusions}
In general the link between default probabilities and credit
spreads is best described by intensity models. The credit spread
to be added to the risk free rate represents a good measure of a
bond credit risk for example. Yet, intensity models present some
drawbacks: They do not link the default event to the economy but
rather to an exogenous jump process whose jump component remains
economically unexplained. 
Also, when dealing with default correlation, a copula function
must be introduced between the jump processes thresholds in a way
that has no clear immediate relation with equity correlation, a
source of correlation that can be used for practical purposes.

In this paper we introduced an analytically tractable structural
first passage model (SVBAT1P) based on scenarios on the value of
the firm volatility and on the default barrier. This model allows
for a solution to the above points. In this model the default has
an economic cause, in that it is caused by the value of the firm
hitting the default safety barrier value, and all quantities have
a clear economic interpretation. Also, when dealing with
multi-name products, the model allows for the introduction of the
correlation in a very natural way, by simply correlating shocks in
the different values of the firms by means of equity correlation.


We showed how to calibrate the model parameters to actual market
data: Starting from CDS quotes and the scenario based barrier
model, we first devised a calibration method based on the kernel
of a matrix built by CDS with different maturities and under
different scenarios. This approach involves a model with more
parameters than market quotes, so that part of the parameters have
to be fixed arbitrarily. After showing the limits of this
approach, we considered a number of parameters equal to the number
of quotes and calibrated via an optimization. This approach,
examined in a case study, pointed out that the scenario based
barrier parameterization is not describing well the implicit
dynamics and debt structure given in CDS quotes for different
maturities. An improvement is obtained when allowing for different
scenarios also on the firm value volatility and also when
introducing weights in the objective function, leading to a
calibration error essentially within the bid ask spread. Pricing
of counterparty risk in equity payoffs shows a partial consistency
with the deterministic coefficients AT1P model in Brigo and
Tarenghi (2004) calibrated to the same data.

Finally, the S(V)BAT1P model can be used to build a relationship
between the firm value $V$ and the firm equity $S$ (perceived as a
suitable barrier payoff in terms of $V$ itself), for example along
the lines of Jones et al (1984) and Hull, Nelken and White (2004).
This approach can be used to price an equity default swap. To do
so we need to find an expression for the debt (and thus the
equity) within the chosen structural model. Debt and equity
expressions are known in closed form for time-constant and
standard (exponential) barrier Black Cox models (see for example
Bielecki and Rutkowski (2001), Chapter 3). Under SVBAT1P we simply
obtain a linear combination of said formulas under each scenario
and we have a closed form expression for the equity. Then we can
price an equity default swap by means of Monte Carlo simulation of
the firm value, from which, scenario by scenario, we deduce
analytically the equity value by means of the found formula. This
is currently under investigation.

\end{document}